\documentclass[prb,aps,showpacs,preprint,superscriptaddress]{revtex4}
\usepackage{graphics}
\usepackage{amsmath}
\usepackage{amsfonts}
\usepackage{amssymb}
\usepackage{graphicx}
\usepackage{ulem}
\usepackage{soul}
\usepackage{color}
\begin{document}
\title{Electronic Structure and magnetism in Ir based double-perovskite Sr$_2$CeIrO$_6$}

\author{S. K. Panda}
\affiliation{Centre for Advanced Materials, Indian Association for the Cultivation of Science, Jadavpur, Kolkata, 700032, India.}
\author{I. Dasgupta}
\affiliation{Centre for Advanced Materials, Indian Association for the Cultivation of Science, Jadavpur, Kolkata, 700032, India.}
\affiliation{Department of Solid State Physics, Indian Association for the Cultivation of Science, Jadavpur, Kolkata, 700032, India}
\begin{abstract}
The electronic structure and magnetism of Sr$_2$CeIrO$_6$, an Ir-based double perovskite system has been investigated using first-principles calculations. 
We found that a strong spin-orbit coupling dictate the electronic and magnetic properties of this system. A small value of U along with SOC could
open up a gap at the Fermi level, offering the possibility of novel J$_{eff}$ = $\frac{1}{2}$ Mott quantum state. Our calculations reveal that the magnetic
ground state is antiferromagnetic in agreement with the magnetization data and provide the value of spin and orbital moment for this system which is
in agreement with the other isostructural Ir-based compound.
\end{abstract}
\maketitle
\section{Introduction}
Until few years ago, the common belief has been that due to the extended nature of the 5d orbitals, electron correlation does not play any significant role
in 5d transition metal oxides (TMO) and density functional theory (DFT) within local density approximation (LDA) or generalized gradient approximation (GGA) 
can explain the metallic ground state of these systems. However, recently there are reports of insulating ground state in some Ir-based oxides such
as Sr$_2$IrO$_4$~\cite{Sr2IrO4_Insulator}, Na$_2$IrO$_3$~\cite{Na2IrO3}, Sr$_3$Ir$_2$O$_7$~\cite{Sr3Ir2O7} etc. It has been shown that a combined effect of
strong spin-orbit coupling (SOC) and onsite Coulomb correlation ($U$) can lead to an antiferromagnetic insulating ground state in such irridates with half-filled
(5d$^5$) Ir$^{4+}$ ions~\cite{Sr2IrO4_SOC1,Sr2IrO4_SOC2}. Since the interplay of SOC and $U$ dictate the physics of irridates, it has gained lot of attention
in last few years and many exotic phases like quantum spin liquid, topological insulator have been realized in these
systems~\cite{QSL_Na4Ir3O8,QSL_Ba3IrTi2O9,TI_Na2IrO3_1,TI_Na2IrO3_2}. Double perovskite systems
with formula A$_2$BB$^\prime$O$_6$ have attracted substantial attention during the last couple of decades due to many novel properties and
it is therefore likely that a combined effect of SOC and $U$ may lead to interesting properties in such systems. 
Recently within a model Hamiltonian approach,
Chen \textit{et al}~\cite{doublePerovskite_SOC_LeoBalents} have shown the possibility
of exotic phases induced by strong SOC in completely ordered double perovskite A$_2$BB$^\prime$O$_6$ where B$^\prime$ has either 4d$^1$ or 5d$^1$ electronic
configuration. The importance
of spin-orbit coupling in Ir based double perovskites La$_2$CoIrO$_6$ and Sr$_2$CoIrO$_6$ have been studied by Kolchinskaya \textit{et al}~\cite{La2CoIrO6_XMCD}
using x-ray magnetic
circular dichroism (XMCD) measurement and they find that the orbital moment is quite large and almost equal to the spin moment.
In this context, Ir-based double perovskites may be interesting and warrant a detailed theoretical investigation. It would be more interesting
to study a Ir-based double perovskite system where Ir is in 5d$^5$ electronic configuration which can then be a potential candidate for exhibiting novel
J$_{eff}$=$\frac{1}{2}$ Mott ground state. The novel J$_{eff}$=$\frac{1}{2}$ Mott state has already been suggested in many other iridates with half-filled
(5d$^5$) Ir $4+$ ions like Sr$_3$Ir$_2$O$_7$~\cite{Sr3Ir2O7_jHalf}, CaIrO$_3$~\cite{CaIrO3_Alaska},
Sr$_2$IrO$_4$~\cite{Sr2IrO4_SOC1} and this recently discovered quantum state can give rise to various kind of interesting low-energy
Hamiltonians for instance the highly anisotropic Kitaev model which may be important for quantum computing~\cite{Sr2IrO4_SOC2}.  
With this motivation, we chose a well-ordered double perovskite Sr$_2$CeIrO$_6$ where Ir is in 4+ charge state (d$^5$ configuration)
and Ir is the only magnetically active ion in this system.
\begin{figure}
\begin{center}
\includegraphics[scale=0.70]{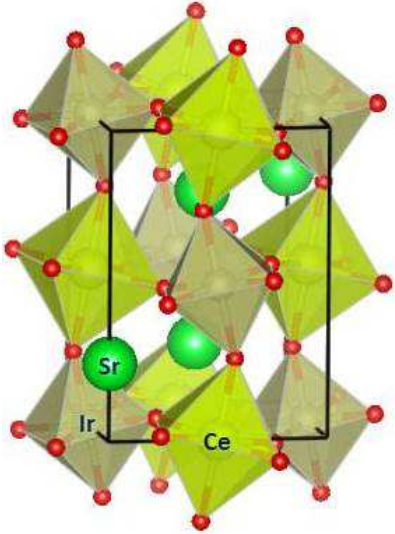}
\end{center}
\caption{Crystal structure of Sr$_2$CeIrO$_6$.}
\label{Cryst_Struct}
\end{figure}
\par
Sr$_2$CeIrO$_6$ crystallize in monoclinic structure and shows an antiferromagnetic transition at 21 K~\cite{Sr2CeIrO6_CrystStruct}.
Neutron powder diffraction (NPD) study~\cite{Sr2CeIrO6_Neutron} did not see any magnetic Bragg peak and hence concluded that the magnetic moment at the Ir site
is very small. NPD measurement of isostructural compound La$_2$CoIrO$_6$~\cite{La2CoIrO6_Neutron} where Ir is in d$^5$ state, also did not give any information of
the magnetic ordering at the Ir site.
However XMCD measurement~\cite{La2CoIrO6_XMCD}
on La$_2$CoIrO$_6$ clearly see substantial magnetization at the Ir site and the estimated spin and orbital moments are respectively 0.20 $\mu_B$ and 0.18 $\mu_B$ per Ir.
This difference between NPD and XMCD results is attributed to the low neutron scattering cross section for Ir. 
Since neutron
diffraction study did not provide any insight to the magnetic ordering and no XMCD measurements are available for Sr$_2$CeIrO$_6$, a theoretical investigation
will therefore be very useful to determine the type of magnetic ground state in Sr$_2$CeIrO$_6$. In view of the above, we have investigated the magnetic properties
of Sr$_2$CeIrO$_6$
within the density functional approach and find the system to be spin-orbit coupling driven Mott-Hubbard antiferromagnetic insulator where a
novel J$_{eff}$=$\frac{1}{2}$ state can be realized. The remainder of the paper is organized as follows. In section~\ref{comp} we shall describe the 
computational details and the crystal structure of Sr$_2$CeIrO$_6$. Finally section~\ref{result} will be devoted to the presentation and discussion of the
results of our calculations.
\begin{figure}
\begin{center}
\includegraphics[scale=0.5]{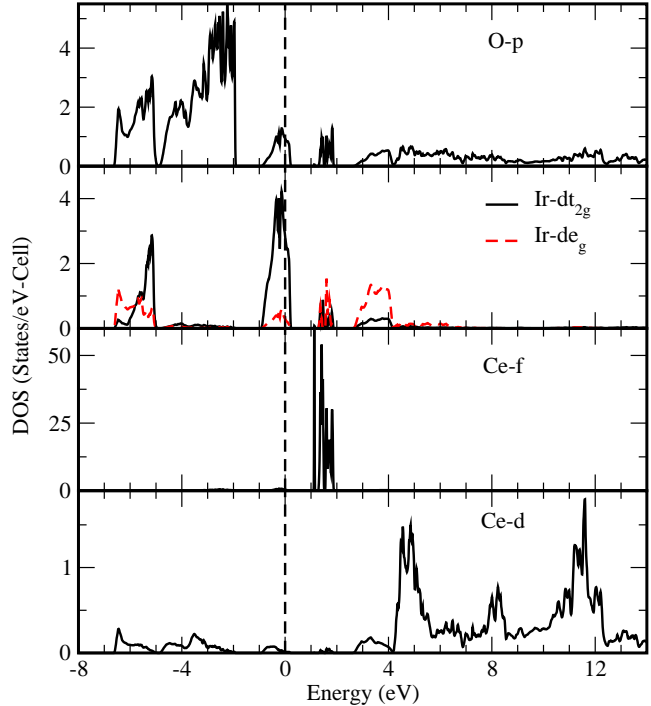}
\end{center}
\caption{Partial Density of States for non-spin polarized configuration.}
\label{dos_nm}
\end{figure}
\section{Computational detail and Crystal Structure}
\label{comp}
All the calculations reported in this work are carried out using full potential linearized augmented plane wave~(FPLAPW) method as embodied in
WIEN2k code~\cite{wien2k}. The atomic sphere radii($R_{MT}$) of Sr, Ce, Ir and O are chosen
to be 2.37~a.u., 2.22~a.u., 2.07~a.u., and 1.69~a.u. respectively. To achieve energy convergence
of eigenvalues, the wave functions in the interstitial region were expanded in plane waves with a cut-off $R_{MT}k_{max}$=7, where $R_{MT}$ denotes the smallest
atomic sphere radius and $k_{max}$ represents the magnitude of the largest k vector in the plane wave expansion. The valence wave functions inside the spheres
are expanded up to $l_{max}$=10 while the charge density was Fourier expanded up to $G_{max}$=12. The Brillouin-zone integration was done with a modified tetrahedron
method~\cite{tetrahedron} using 130 special k points in the irreducible part of the Brillouin-zone to achieve self-consistency. Exchange and correlation effects
are treated within the DFT using LSDA with parameters of Moruzzi, Janak and Williams~\cite{LSDAPARA} and a Hubbard U was included in
the framework of LSDA+U~\cite{LSDAU_Localized1,LSDAU_Localized2}. A moderate value
for the effective on-site Coulomb interaction U$_{eff}$ = 2.0 eV is taken to describe the effect of correlation. The effect of spin-orbit coupling was
treated using a second-variational scheme. 
\begin{table}
\caption{Energy of FM and AFM state along with the spin and orbital moments. Energy for FM state is assumed to be zero. For the LSDA+U+SOC
calculations, orbitals moment of Ir is written in the parenthesis.}
\centering
\begin{tabular}{|c| c| c| c|c|c|c|}
\hline
 & & Energy &\multicolumn{4}{|c|}{Spin (Orb) Moment} \\
 & & (meV/fu) &\multicolumn{4}{|c|}{($\mu_B$)} \\[1 ex]
 & & & Ir & O & Int & Tot \\
\hline
& FM & 0.0 & 0.57 & 0.06 & 0.21 & 2.0 \\[-1ex]
\raisebox{1.5ex}{LSDA+U} &AFM
& 27.21 & 0.44 & 0.0 & 0.0 & 0.0 \\[1ex]
\hline
& FM & 0.0 & 0.44 (0.40) & 0.04 & 0.21 & 1.53 \\[-1ex]
\raisebox{1.5ex}{LSDA+U+SOC} &AFM
& -9.72 & 0.40 (0.42) & 0.0 & 0.0 & 0.0 \\[1ex]
\hline
\end{tabular}
\label{table:energy}
\end{table}
\par
Sr$_2$CeIrO$_6$ crystallize with a monoclinic symmetry in the space group $P2_1/n$. The deviation from the ideal double perovskite structure is the tilting of
the corner-sharing CeO$_6$ and IrO$_6$ octahedra which break the cubic symmetry and also the elongation of the Ce-O bond length to 2.2~\r{A}
at the expense of Ir-O bond length which become 2.0~\r{A}. The tilting of the two octahedra in opposite direction make the Ir-O-Ce angle
to be $156^{\circ}$. There is a little distortion in the octahedra in terms of angles but the bond-length with O ions are all equal. The crystal structure is shown
in Fig.~\ref{Cryst_Struct}. Unit cell contain two formula units. Coordinates of the two equivalent Ir, located at 2c site are (0.0, 0.5, 0.0) and (0.5, 0.0, 0.5).
Since the distance between this two Ir is quite large (5.82~\r{A}), they could only interact via O-Ce-O path. All the
calculations are performed with the experimental atomic positions and lattice
parameters a = 5.8180~\r{A}, b = 5.8402~\r{A}, c = 8.2355~\r{A}, and $\beta$ = $90.225^{\circ}$~\cite{Sr2CeIrO6_Neutron}.

\begin{figure}
\begin{center}
\includegraphics[scale=0.5]{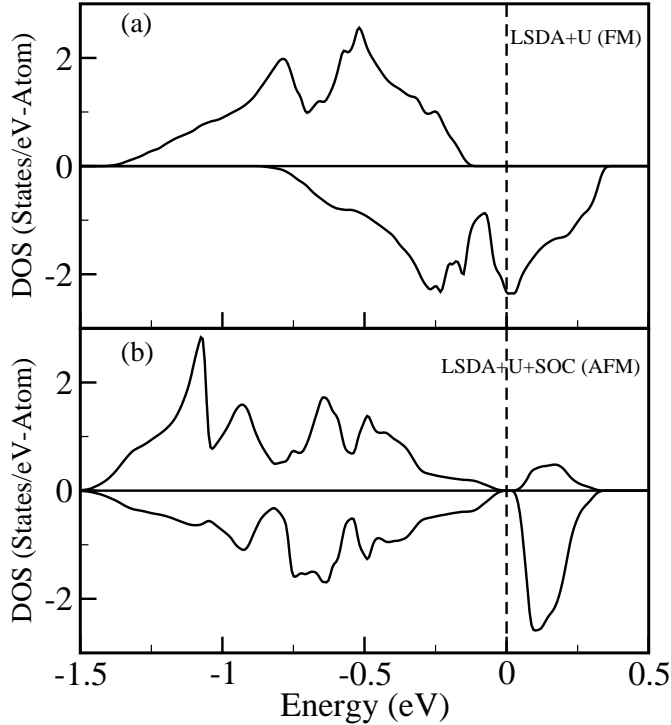}
\end{center}
\caption{Partial Density of States of Ir-d of both the spin channels for (a) FM ground state within LSDA+U and (b) AFM ground state within LSDA+U+SOC.}
\label{dos_afm}
\end{figure}
\section{Results and Discussion}
\label{result}
We have analyzed the electronic structure of non-spin polarized LDA calculations to get an estimate of the charge state of Ce and Ir ions and 
the octahedral crystal field splitting.
The calculated partial density of states (Fig.~\ref{dos_nm}) shows that both Ce-d and Ce-f derived states are completely empty.
This confirm that in the ionic limit, electronic structure is consistent with the $4+$ charge state of Ce in agreement with the experiment~\cite{Sr2CeIrO6_Neutron}
and hence Ce will be magnetically inactive in this system.
Ir is therefore expected to be in $4+$ (d$^5$) charge state. The partial density of states (PDOS) of Ir-d (Fig.~\ref{dos_nm})
reveal that the Ir-d $e_g$ ($x^2-y^2$ and $3z^2-1$) states are empty while the $t_{2g}$ ($xy$, $xz$ and $yz$) states are partially occupied
consistent with the $4+$ charge state of Ir and the crystal field splitting between $t_{2g}$ and $e_g$ state is appreciable ($\sim$2.5 eV). Further there is
substantial hybridization between Ir-d and O-p states.   
Since IrO$_6$ octahedra are distorted and also tilted by an angle of $12^{\circ}$ from the c-axis,
this can lead to some admixture between $t_{2g}$ and $e_g$ states. However Fig.~\ref{dos_nm} clearly reveal that the mixing between
$t_{2g}$ and $e_g$ states
are very little. Since $t_{2g}$ and $e_g$ states are separated by a large gap (2.5 eV) and they hardly mix, one can assign an effective 
angular quantum number $l_{eff}=1$ for the $t_{2g}$ manifold as in Sr$_2$IrO$_4$~\cite{Sr2IrO4_SOC1}. We also note that the tilting angle
of IrO$_6$ octahedra in Sr$_2$IrO$_4$ is $11^{\circ}$~\cite{Sr2IrO4_CrystStruct}.

\begin{figure}
\includegraphics[width=\columnwidth]{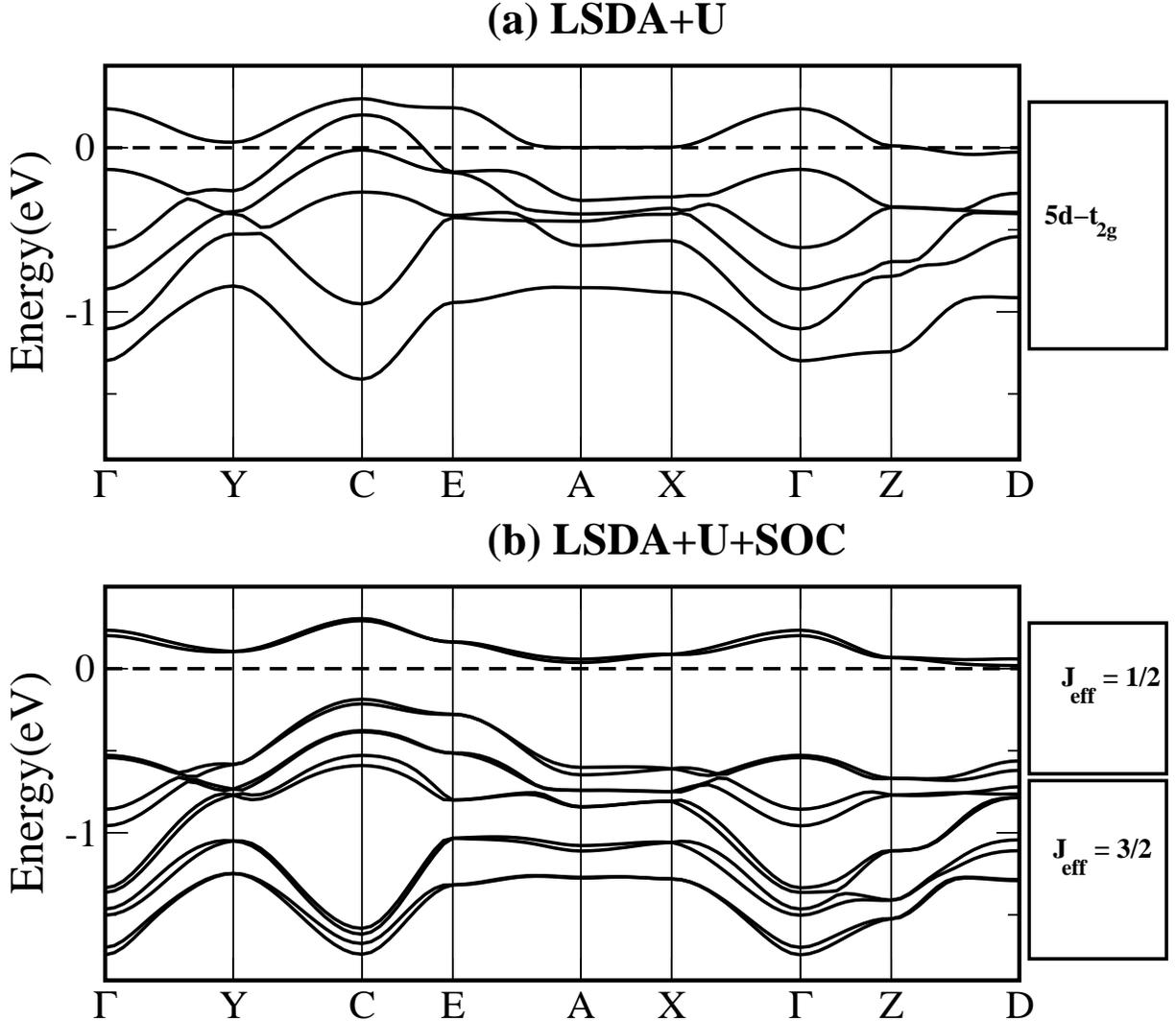}
\caption{Band dispersion along various high symmetry directions in the AFM state, calculated within (a) LSDA+U and (b) LSDA+U+SOC approach.}
\label{Bands_AFM}
\end{figure}
\par
We have next considered the magnetic properties of Sr$_2$CeIrO$_6$. Our LDA calculations reveal that the magnetic solution can only be stabilized in the presence
of Hubbard $U$ at the Ir site. The results of our calculations for the ferromagnetic and antiferromagnetic arrangement of Ir spins are summarized in
Table.~\ref{table:energy}. From Table~\ref{table:energy} we gather that the half metallic ferromagnetic state (see Fig.~\ref{dos_afm}-(a))
with integral magnetic moment is stable when the
calculations are carried out in the framework of LSDA+U method. This can be easily understood as in the spin polarized calculation, the $t_{2g}$ states in the
majority spin channel are completely occupied while it is partially filled in the minority spin channel leading to a stable half metallic state as illustrated
in Fig~\ref{dos_afm}-(a). However upon
inclusion of SOC not only the system become insulating (see Fig.~\ref{dos_afm}-(b)) but the AFM state become the stable solution in agreement with the experiment. 
Table~\ref{table:energy} reveal that the spin and orbital
moment are almost equal in magnitude and parallel to each other, establishing the importance of SOC as in other irridates.
The insulating gap is calculated to be 30 meV.
It is interesting to note that recent XMCD measurement~\cite{La2CoIrO6_XMCD} for isostructural
compound La$_2$CoIrO$_6$ also see an equal spin and orbital moment along the same direction at the Ir site. 
\par
In the following we argue that the importance of SOC and insulating solution can be rationalized by invoking J$_{eff}$=$\frac{1}{2}$ state. In Fig.~\ref{Bands_AFM}
we show the LSDA+U and LSDA+U+SOC band dispersion for the AFM ground state. 
The LSDA+U band dispersion near Fermi
level displayed in Fig.~\ref{Bands_AFM}-(a) exhibit six bands from two Ir atoms in the unit cell, arising from the $t_{2g}$ manifold. 
As expected for the d$^5$ configuration of Ir, in the majority spin channel the $t_{2g}$ states of one Ir will be completely occupied while for the other it
will be 2/3 occupied leading to a metallic AFM state. 
A significant change is observed in the band-dispersion upon inclusion of SOC (see Fig.~\ref{Bands_AFM}-(b)) where we have a manifold of twelve SOC bands
arising from the $t_{2g}$ up and dn states of two Ir. 
In the limit of strong SOC the $t_{2g}$ orbitals with an effective quantum state $l_{eff}=1$ form J$_{eff}$ = $\frac{3}{2}$
quartet and J$_{eff}$ = $\frac{1}{2}$ doublet. Here J$_{eff}$ = $\frac{1}{2}$ doublet would be energetically higher than the J$_{eff}$ = $\frac{3}{2}$ bands, since
J$_{eff}$ = $\frac{1}{2}$ is produced from the J$_{5/2}$ manifold due to large crystal field splitting.
Therefore we clearly see in Fig.~\ref{Bands_AFM}-(b) that a branch of total eight bands of two Ir which form J$_{eff}$ = $\frac{3}{2}$ manifold
while the remaining four bands belong to J$_{eff}$ = $\frac{1}{2}$ manifold. Among the total 10 Ir-d electrons in the unit cell, 8 electrons
completely occupy the J$_{eff}$ = $\frac{3}{2}$ bands and hence J$_{eff}$ = $\frac{1}{2}$ bands are half-filled. Therefore a small but finite $U$ can easily
split these half-filled band into upper and lower Hubbard bands, yielding a Mott insulating state (see Fig.~\ref{Bands_AFM}-(b)).
We note that such a mechanism was suggested for Sr$_2$IrO$_4$~\cite{Sr2IrO4_SOC1} as well as CaIrO$_3$~\cite{CaIrO3_Alaska} where J$_{eff}$ = $\frac{1}{2}$
state was realized. We however note that the spin and orbital moments
are calculated to be nearly equal in contrary to the expectation from a $J_{eff}=\frac{1}{2}$ state where the orbital moment is expected to be twice to the spin moment.
The deviation may arise due to the tilting and the distortion of the octahedra as well as because of the Ir-d and O-p hybridization. 
\par
In summary, our calculations find AFM ground state of Sr$_2$CeIrO$_6$ and establish that strong SOC dictate both the electronic and magnetic
properties of this system. We also report the possible manifestation of a novel J$_{eff}$ = $\frac{1}{2}$ Mott state in a Ir-based double perovskite system.


\end{document}